\documentclass[twocolumn,nofootinbib,superscriptaddress,preprintnumbers]{revtex4-1}

\pdfoutput=1

\usepackage{graphicx}

\usepackage{slashed}

\def\be{\begin{equation}}
\def\ee{\end{equation}}

\newcommand{\bea}{\begin{eqnarray}}
\newcommand{\eea}{\end{eqnarray}}

\usepackage{bbm,bm,amsmath,amssymb}

\usepackage{hyperref}

\def\Ls{\Lambda_{\rm s}}

\begin{document}

\preprint{MPP-2024-7}

\title{Species Scale and Primordial Gravitational Waves}

\author{Marco Scalisi}
\email{mscalisi@mpp.mpg.de}
\affiliation{Max-Planck-Institut f\"ur Physik (Werner-Heisenberg-Institut), F\"ohringer Ring 6, 80805 M\"unchen, Germany}

 \begin{abstract}
\noindent
The species scale is  a field-dependent UV cut-off for any effective field theory weakly coupled to gravity.
In this letter we show that, in the context of inflationary cosmology, a detection of primordial gravitational waves will set an upper bound on the decay rate $|\Lambda'_s/\Lambda_s|$ of the species scale. Specifically, we derive this in terms of the tensor-to-scalar ratio $r$ of power spectra of primordial perturbations. Given the targets of  current and next generation experiments, we show that any successful detection would signify that this upper limit is of the order of unity, which is consistent with recent discussions in the literature.

  \end{abstract}

\maketitle

\vspace{1cm}

\section{Introduction}
Common lore suggests that effects of quantum gravity are difficult to unravel.
One of the reasons is that the energy at which one expects quantum gravitational effects to become relevant is the Planck scale \mbox{$M_{\rm P}\simeq 2.48 \times 10^{18}$ GeV}. This is about 14 orders of magnitude above the highest energy reached on Earth, namely at the Large Hadron Collider. 

In recent years, we have however learned that, for gravitational effective field theories (EFTs) with a large number $N_{\rm s}$ of light species, a more natural UV cut-off is the {\it species scale} \cite{Dvali:2007hz,Dvali:2007wp,Dvali:2009ks,Dvali:2010vm,Dvali:2012uq} (see also \cite{Veneziano:2001ah,Arkani-Hamed:2005zuc,Distler:2005hi} for earlier works), a renormalization of the Planck mass, which is given by\footnote{In $d$ dimensions, this formula has a simple generalization via the substitution $\sqrt{N_{\rm s}}\rightarrow N_{\rm s}^{\frac{1}{d-2}}$ and by considering $M_{\rm P}$ as the $d$-dimensional reduced Planck mass.}
\begin{equation}\label{SpeciesScale}
\Lambda_{\rm s}=\frac{M_{\rm P}}{\sqrt{N_{\rm s}}}\,.
\end{equation}
 Above the species scale, gravity becomes strongly coupled and no effective description, with such a number of light particles, is thus possible. In the limit of large number of species, $\Ls$ will significantly deviate from $M_{\rm P}$ so that quantum gravity effects can become important for effective field theories even at lower energies.

In the context of string theory, towers of light particles appear naturally in certain corners. A typical situation is at the boundary of moduli space. Here one can rigorously test the Swampland Distance Conjecture (SDC) \cite{Ooguri:2006in}, which states that the limit of infinite scalar field variations is always associated with an infinite tower of states exponentially becoming light. Analogous conclusions can be drawn in the more conservative case of super-Planckian field excursions \cite{Baume:2016psm,Klaewer:2016kiy}, namely larger that $M_{\rm P}$. In the limit of large distances, the mass decay of the tower determines in fact a drop-off of the species scale, following eq.~\eqref{SpeciesScale}, since the number of light species increases. This simple argument already suggests that the species scale must be a {\it field-dependent} UV cut-off (more evidence about the moduli-dependence of $\Lambda_{\rm s}$ was recently given in \cite{vandeHeisteeg:2022btw,Castellano:2022bvr,Cribiori:2022nke,vandeHeisteeg:2023ubh,Andriot:2023isc,Blumenhagen:2023yws,Cribiori:2023sch,Calderon-Infante:2023ler,vandeHeisteeg:2023dlw,Castellano:2023aum}).

The implications of such species scale drop-off for {\it inflation} have first been studied in \cite{Scalisi:2018eaz} (see also \cite{Scalisi:2019gfv}), where it was shown that a universal upper bound on the inflaton range can be derived (in \cite{vandeHeisteeg:2023uxj} the same bound was pointed out). It was shown that inflationary models, with field ranges violating this bound, do not admit a gravitational weakly coupled effective description, since their typical energy scale is above $\Lambda_{\rm s}$.

The rate at which the species scale varies in field space, namely $\lambda\equiv|\Ls'/\Ls|$, is unknown and depends on the string effective model at hand. It is determined by the decay rate of the tower of states and has been subject of some research activity, mainly aiming at identifying lower bounds for it \cite{vandeHeisteeg:2023ubh,Calderon-Infante:2023ler,Castellano:2023stg,Castellano:2023jjt}. The significance of a lower bound for such decay rate is directly connected to the validity of the SDC. Arbitrarily low values of $\lambda$ would in fact correspond to arbitrarily large distances in field space,\footnote{This argument can also be used to constrain the non-geodicity of a trajectory in field space \cite{Freigang:2023ogu}. In an accelerating background, it has been shown that this must be negligible.} which are forbidden by the SDC. Just more recently, it was suggested \cite{vandeHeisteeg:2023ubh,Calderon-Infante:2023ler,vandeHeisteeg:2023dlw,Lust:2023zql} that the rate $\lambda$ be also upper bounded by an ${\cal O}(1)$ quantity. Namely, the species scale $\Ls$ cannot decrease (or increase) arbitrarily fast in field space. One of the implications is that certain finite scalar field variations must be allowed in string theory.

The prevailing research approach so far has been to study properties of the species scale in the context of well controlled effective string models (see e.g. \cite{vandeHeisteeg:2022btw,Cribiori:2022nke,vandeHeisteeg:2023dlw,Cribiori:2023sch}) and then discuss the potential implications for phenomenology. However, a  risk of this approach is the potential oversight of critical details that a more realistic model may incorporate. One common simplification in these investigations is the assumption of isotropy in Kaluza-Klein directions. Nevertheless, several more realistic models opt for the utilization of anisotropic compactification spaces, such as warped throats, as a means to maintain control over supersymmetry breaking.

In this work, we take the opposite approach, namely we show how (cosmological) observations can provide precious information about properties of the species scale.
Current and upcoming Cosmic Microwave Background (CMB) experiments are going to severely constraint \mbox{$r\equiv A_t/A_s$}, the ratio of amplitutes of tensor and scalar primordial perturbations. The current upper bound is $r<0.036$ \cite{BICEP:2021xfz}. Next generation experiments, such as  CMB-S4 and the Simons Observatory \cite{CMB-S4:2016ple,SimonsObservatory:2018koc}, will target smaller values, namely $r=\mathcal{O}(10^{-3})$. Any detection would be an important indication that  {\it primordial gravitational waves} were generated during inflation.

As key result of this letter, we demonstrate that any future detection of the tensor-to-scalar ratio $r$ will set a universal upper bound on the species scale decay rate as
\begin{equation}\label{upperboundmain}
    \lambda=\left|\frac{\Ls'}{\Ls}\right|\lesssim  \frac{c}{\sqrt{r}}\log\frac{10^8}{r}
\end{equation}
with $c=\mathcal{O}(10^{-2}$--$10^{-3})$ depending on the specific class of inflation models. As universal and most conservative estimate, we find that $c<\sqrt{8}/2N$, with $N$ being the total number of e-foldings of quasi-exponential expansion. We show how the constant $c$ changes and is strengthened in a variety of generic models of inflation. In particular, we study the cases of chaotic monomial, Starobinsky-like, hilltop and inverse-hilltop (or brane inflation) potentials. As key-result, we find that $\lambda$ is upper bounded by an {\it order one quantity}. This is in agreement with recent theoretical arguments given in \cite{vandeHeisteeg:2023ubh,Calderon-Infante:2023ler,vandeHeisteeg:2023dlw,Lust:2023zql}.

The outline of this work is as follows: we first review some of the relevant aspects and formulas of the species scale (in sec.~\ref{sec:speciesscale}) and of inflation and the associated scalar field variation (in sec.~\ref{sec:inflation}). In sec.~\ref{sec:bounds}, we present our main results. In sec.~\ref{sec:conclusions}, we draw our conclusions and offer an outlook for such a research direction.

\section{Field-dependent Species scale}\label{sec:speciesscale}

The species scale was introduced to account for modifications of scale hierarchies in gravitational effective field theories with large number of species. There are at least two different ways to understand its importance:
\begin{itemize}
    \item First, $\Lambda_s$ can be defined as the scale at which gravity becomes strongly coupled and perturbation theory breaks down. In a pertubative expansion of an EFT with a large number $N_{\rm s}$ of species weakly coupled to gravity, $\Ls$ is in fact the scale at which the one loop term is of the same order of the tree level term. This happens when the graviton propagator diverges, as it was shown in \cite{Dvali:2007wp}. In this limit, higher derivative terms cannot be neglected \cite{vandeHeisteeg:2023ubh,Cribiori:2023ffn,Calderon-Infante:2023uhz,vandeHeisteeg:2023dlw,Basile:2023blg}.

    \item  A second, perhaps more intuitive, argument is  based on black holes physics \cite{Dvali:2007hz,Dvali:2007wp} (see also \cite{Basile:2024dqq}). In this case, $\Ls$ is defined as the scale corresponding to the smallest black hole still consistent with the EFT. Black holes of Planckian size, with $R_{\rm BH}\simeq M^{-1}_{\rm P}$, would in fact correspond to entropies of order one, using the standard Bekenstein-Hawking formula \mbox{$S_{\rm BH}\simeq R_{\rm BH}^{d-2} M_{\rm P}^{d-2}$}, in $d$ dimensions. However, this goes against the natural expectation for the entropy to be proportional at least to the number $N_{\rm s}$ of light species. One can verify that the minimum size for a black hole to solve this conundrum is precisely $R_{\rm min}=N_{\rm s}^{\frac{1}{d-2}}M_{\rm P}^{-1}$, which is nothing but $\Ls^{-1}$. 
\end{itemize}
Therefore, both arguments lead to the formula eq.~\eqref{SpeciesScale} in 4 dimensions, as given in the Introduction.

In string theory, one expects all parameters to be dependent on vev of moduli or scalar fields. The number of light species $N_{\rm s}$ is no exception. In fact, it has been recently pointed out that, in type II compactifications, one can identify $N_{\rm s}$ with the topological string free energy \cite{vandeHeisteeg:2022btw,Cribiori:2022nke} and, therefore, extract an explicit moduli-dependence of the species scale.

For large distances in field space,\footnote{We define large field variations as $\Delta\phi\gtrsim M_{\rm P}$. This can definitely happen towards the boundary of moduli space, where one can even probe infinite scalar field displacements. In the bulk, one can also traverse modest super-Planckian field ranges.} namely in the regime where one can successfully apply the SDC, the species scale exhibits a typical exponential dependence in terms of the field distance $\Delta\phi$. This can be understood by considering that the number of light species is expected to increase inversely proportional to the typical mass scale as
\begin{equation}
    N_{\rm s}\sim m^{-\alpha} M_{\rm P}^{\alpha}\,,
\end{equation}
with $\alpha$ some order one number. This implies that 
\begin{equation}
    \Ls\sim m^{\frac{\alpha}{2}}M_{\rm P}^{1-\frac{\alpha}{2}}\,,
\end{equation}
using eq.~\eqref{SpeciesScale}. Since the SDC predicts that the tower mass exponentially decreases in field space as $m\sim e^{-\gamma \Delta\phi}$, then one obtains
\begin{equation}\label{speciesscaledecay}
    \Ls=\Lambda_0\ e^{-\lambda \Delta \phi}\,,
\end{equation}
where $\Lambda_0\leq M_{\rm P}$ is the value of the species scale corresponding to zero field displacement and $\lambda=\alpha \gamma /2$ corresponds to the decay rate of the species scale, as given in the Introduction of this article.
As a consequence of eq.~\eqref{speciesscaledecay}, one obtains a universal upper bound on the scalar field range, such as
\begin{equation}\label{DeltaUpper}
   \Delta\phi\leq \frac{1}{\lambda} \log\frac{M_{\rm P}}{\Ls}\,.
\end{equation}
The latter equation implies that certain finite scalar field variations in quantum gravity are allowed at the cost of decreasing the cut-off from the Planck scale. In fact, in the  limit $\Lambda_s\rightarrow M_{\rm P}$, no field displacement is permitted.  Conversely, infinite scalar field variations are prohibited in quantum gravity since the limit $\Lambda_s\rightarrow 0$ corresponds to a complete breakdown of the EFT.

\section{Inflation and Field-range}\label{sec:inflation}

Cosmic inflation is a primary example of an effective field theory, which requires a certain finite scalar field variation $\Delta\phi$ to be successfully realized. The magnitude of this range is strictly model-dependent and can be obtained by the integral
\begin{equation}
   \Delta\phi=\int \sqrt{2 \epsilon}\ {\rm d} N
\end{equation}
calculated over the total period of inflation. In the above equation, $\epsilon$ is the acceleration parameter defined as \mbox{$\epsilon\equiv {\rm d} \log H/{\rm d} N \leq 1$}, with $H$ being the Hubble parameter and $N$ the number of inflationary e-foldings, defined by ${\rm d}N\equiv - H {\rm d}t$. Knowing the exact functional form of $\epsilon(N)$ allows to have a precise estimate of the inflaton range $\Delta\phi$. From the theoretical viewpoint, this can be obtained once we have the explicit form of an inflationary scalar potential. It is instead more challenging to infer the precise form of $\epsilon(N)$ just from observations. One of the reasons is that CMB measurements do not allow us to have access to the whole inflationary trajectory\footnote{This is mainly due to the suppression of the CMB power spectrum at small
angular scales.}, rather just to a restricted window of around 8 e-foldings. The result is that only a small part of the area under the curve $\sqrt{2\epsilon(N)}$ can be determined.

A generic lower bound for $\Delta\phi$ was pointed out by Lyth in his seminal paper \cite{Lyth:1996im}.  Assuming that $\epsilon(N)$ is a monotonic increasing (towards the end of inflation) function of $N$, one can extract the lower bound \cite{Boubekeur:2005zm,Garcia-Bellido:2014wfa}
\begin{equation}\label{Lyth}
   \Delta\phi \gtrsim \sqrt{\frac{r}{0.002}} \,,
\end{equation}
where $r$ is the tensor-to-scalar ratio and is related to the acceleration parameter as $r=16\,\epsilon$, calculated at the moment of horizon exit. Moreover, in the previous formula, we have assumed a total of 60 e-foldings of inflationary expansion. The significance of eq.~\eqref{Lyth} is that, given the target of current or upcoming experiments being $r=\mathcal{O}(10^{-3})$, any future detection of $r$ will point at inflationary EFTs with Planckian or super-Planckian field displacements. This motivates the study of the SDC in this context.

An interesting and very generic class of inflationary models is characterized by the acceleration parameter scaling as \cite{Mukhanov:2013tua,Roest:2013fha,Garcia-Bellido:2014eva,Garcia-Bellido:2014wfa}
\begin{equation}\label{epsilon}
   \epsilon = \frac{\beta}{N^p} \,,
\end{equation}
with $\beta$ and $p$ being some constants. In the latter equation, we display just the leading term and neglect higher order terms in $1/N$. This is justified given the large total number of e-foldings $N\sim 60$ required to explain homogeneity and isotropy of the CMB. In this class of models, we find some of the most studied inflationary scenarios, such as chaotic monomial potentials ($p=1$), inverse-hilltop models ($1 < p < 2$), Starobinsky-like inflation ($p = 2$) and hilltop potentials
($p > 2$).

For this class of models, it has been shown that the inflaton field range $\Delta\phi$ exhibits certain universal properties in the large-$N$ limit and one can obtain a well-approximated estimate of it. In particular, for any $p < 2$ the field range \cite{Garcia-Bellido:2014wfa} is given by
\begin{equation}\label{eq:Deltapl2}
\Delta\phi \simeq \frac{2\sqrt{2\beta}}{2-p} N^{1-\frac{p}{2}}\,,
\end{equation}
where we have neglected the sub-leading contributions related to the point in field space where inflation terminates. As an example, quadratic inflation ($V\sim\phi^2$) is given by $\beta=1/2$ and $p=1$ and, in fact, returns $r=8/N\simeq0.13$ and $\Delta\phi\simeq 2 \sqrt{N}\simeq 15$ for $N=60$.

For $p=2$, the leading part of the scalar field range is given by a
\begin{equation}
\Delta\phi \simeq \sqrt{2\beta}\log{N}\,.
\end{equation}
One example of this class is given by the famous Starobinsky model \cite{Starobinsky:1980te}, which corresponds to $\beta=3/4$ so that it gives $r= 12/N^{2}\simeq0.003$ and $\Delta\phi\simeq\sqrt{3/2}\log N\simeq 5$ for $N=60$. The more general class of $\alpha$-attractors models \cite{Kallosh:2013yoa,Roest:2015qya,Scalisi:2015qga} corresponds instead to $\beta=3\alpha/4$ and, thus, $r= 12 \alpha/N^{2}$ and $\Delta\phi\simeq\sqrt{3\alpha/2}\log N$.

For $p>2$, the $N$-dependent piece of the scalar field range will be sub-leading, namely it will be proportional to a negative power of $N$ (see eq.~\eqref{eq:Deltapl2}). The leading part will be instead corresponding to the point where inflation ends. The benchmark class of models, in this case, is the one of hilltop scalar potentials and any small deviation from it, such as
\begin{equation}\label{hilltop}
V(\phi)= V_0\left[1-\left(\frac{\phi}{\mu}\right)^n+\sum_{q=n+ 1}^{+ \infty}c_q \left(\frac{\phi}{\mu}\right)^q\right]\,,
\end{equation}
where $n$ is a positive number related to the power $p$ as $n=2(1-p)/(2-p)$, the coefficients $c_q$ parametrize deviations from hilltop models. It was shown in \cite{Garcia-Bellido:2014wfa} that in the small-$\mu$ limit (and in the large-$N$ limit) one can still obtain universal predictions for the CMB cosmological observables, in particular
\begin{equation}
r= 2^{5-2p}\frac{(p-2)^{2p-2}}{(p-1)^{p-2}}\frac{\mu^{2p-2}}{N^p}\,,
\end{equation}
and for the inflaton range
\begin{equation}
\Delta \phi\simeq \left[\frac{p-2}{\sqrt{2}(p-1)}\right]^{1-\frac{2}{p}}\mu^{2-\frac{2}{p}}\,,
\end{equation}
where the coefficient $c_q$ enter only sub-leading terms. Note that the leading term of $\Delta\phi$ does not depend on $N$, as already announced above.

\section{Bounds on species scale}\label{sec:bounds}

Now, we are ready to discuss the universal upper bound on the species scale decay rate. We will show that focusing on specific classes of inflation models, as presented above, allows to strengthen the bound. The starting point is noting that eq.~\eqref{DeltaUpper} can be recast into a bound on the slope of the species scale such as
\begin{equation}
    \left|\frac{\Ls'}{\Ls}\right| \leq \frac{1}{\Delta\phi} \log\frac{M_{\rm P}}{H}\,,
\end{equation}
where we have used that $H\leq\Ls$ for consistent gravitational inflationary EFTs. Then, one can express the Hubble parameter $H$ at CMB scales in terms of the amplitudes of scalar perturbations $A_s$ and the tensor to scalar ratio $r$, such as $H=\sqrt{\pi^2 A_s r/2}~
M_{\rm P}$, and obtain
\begin{equation}
    \left|\frac{\Ls'}{\Ls}\right| \lesssim \frac{1}{2 \Delta\phi} \log\frac{10^8}{r}\,,
\end{equation}
where we have used the measured valued $A_s\approx 2.1\times 10^{-9}$ \cite{Planck:2018vyg}. From the expression above, one can already notice that, for current and near-future targets of $r$ (namely $r\simeq 10^{-2}-10^{-3}$), any scalar field range of order $\mathcal{O}(1-10)$ would correspond to an upper bound on the species scale slope of order $\mathcal{O}(10-1)$, respectively.

\begin{figure}[t]
	\begin{center}
		\includegraphics[scale=0.62]{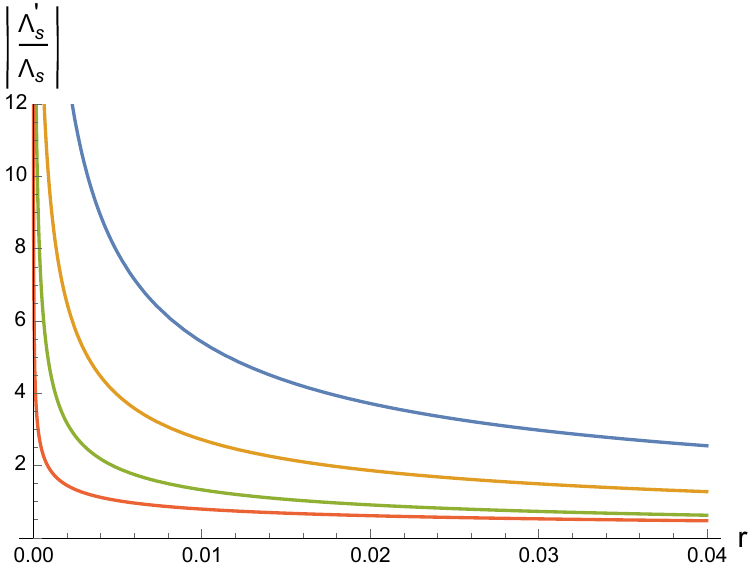}
		\vspace*{-0.2cm}\caption{\it Upper bounds on the species scale decay rate $|\Ls'/\Ls|$. The blue line is the most conservative estimate for an upper bound and it is reproduced by using the Lyth bound for $\Delta\phi$. The other lines represents upper bounds on  $|\Ls'/\Ls|$ for inflation models belonging to the generic class eq.~\eqref{epsilon} with $p=\{1,2,3\}$ respectively.}
		\label{upperbounds}
	\end{center}
\end{figure}

It is possible to obtain a more precise estimate for the upper bound on $|\Ls'/\Ls|$ by expressing the scalar field range $\Delta\phi$ as function of $r$. A first conservative result is given by using the Lyth bound, that is eq.~\eqref{Lyth}. This yields
\begin{equation}
    \left|\frac{\Ls'}{\Ls}\right| \lesssim \frac{1}{30\sqrt{2 r}}\log\frac{10^8}{r}\,. 
\end{equation}
This function is displayed in fig.~\ref{upperbounds} as a blue line.

One can strengthen the bound by considering specific class of models. Chaotic {\it monomial potentials} ($V\sim\phi^n$) have in fact $\Delta\phi \simeq 30 \sqrt{2r}$ (obtained by combining eq.~\eqref{epsilon}  and eq.~\eqref{eq:Deltapl2} for $p=1$ and $N=60$), which strengthens the bound of a factor of 2 and, therefore, yields
\begin{equation}
    \left|\frac{\Ls'}{\Ls}\right| \lesssim \frac{1}{60\sqrt{2 r}}\log\frac{10^8}{r}\,. 
\end{equation}
This function is displayed as a yellow line in fig.~\ref{upperbounds}.

{\it Starobinsky-like potentials} with scalar spectral index $n_s=1-2/N$ (namely corresponding to $p=2$) are characterized by $\Delta\phi\simeq 15\log(60)\sqrt{2r}$, which strengthens again the previous bound such as 
\begin{equation}
    \left|\frac{\Ls'}{\Ls}\right| \lesssim \frac{1}{30\log(60)\sqrt{2 r}}\log\frac{10^8}{r}\,,
\end{equation}
for a total of 60 e-foldings of inflation. This function is displayed as a green line in fig.~\ref{upperbounds}.

For the class of {\it hilltop models} and its deviations, as given by eq.~\eqref{hilltop} (corresponding to $p>2$), one can show that $\Delta\phi\sim r^{\frac{1}{p}}$. By keeping track of the correct coefficients, this leads to 
\begin{equation}
    \left|\frac{\Ls'}{\Ls}\right| \lesssim 2^{\frac{4}{p}-\frac{7}{2}}\frac{p-2}{15}r^{-\frac{1}{p}}\log\frac{10^8}{r}\,,
\end{equation}
again when setting $N=60$. This function for $p=3$ ($n=4$) is displayed as an orange line in fig.~\ref{upperbounds}. One can notice that, within this class of models, a detection of primordial gravitational waves with $r$ close to the current upper bound ($r\lesssim 0.036$), would signify that to have a strong upper bound on the decay rate of the species scale such as $|\Ls'/\Ls|\lesssim 0.5$.

Finally, we discuss the class of {\it inverse hilltop models} or {\it brane inflation}, characterized by a scalar potential of the form
\begin{equation}\label{inversehilltop}
V(\phi)= V_0\left[1-\left(\frac{\mu}{\phi}\right)^n+\ldots\right]\,,
\end{equation}
where the dots represents sub-leading deviations during inflation and $n$ is a positive power, related to \mbox{$1<p<2$}  as \mbox{$n=2(p-1)/(2-p)$}. This class has gained much attention in the context of string and supergravity effective models (see e.g. \cite{Dvali:1998pa,Kachru:2003sx,Kallosh:2022feu}). More recently, the authors of \cite{Burgess:2022nbx} have argued for an improved version of the famous $D3$-$\overline{D3}$ inflation model, which falls again into this class. The inflaton range can be expressed in terms of the tensor-to-scalar ratio $r$ as $\Delta\phi\simeq 30\sqrt{2r}/(2-p)$, which leads to a strengthened bound 
\begin{equation}
    \left|\frac{\Ls'}{\Ls}\right| \lesssim \frac{1}{(2+n)30\sqrt{2 r}}\log\frac{10^8}{r}\,, 
\end{equation}
again when we set $N=60$ and express $p$ in terms of the power $n$. It is interesting to notice that by appropriately choosing the power $n$ in eq.~\eqref{inversehilltop} one can severely lower the upper bound. For example, $n=4$ (typical $D3$-$\overline{D3}$ scenario \cite{Kachru:2003sx,Burgess:2022nbx}) corresponds to an upper bound which is lower than the one obtained for Starobinsky-like potentials ($p=2$). A detection of $r$ close to the current experimental upper bound ($r\lesssim 0.036$), would correspond again to $|\Ls'/\Ls|\lesssim 0.5$.

\vspace{-0.1cm}

\section{Conclusions}\label{sec:conclusions}
In this letter, we have showed that a detection of primordial gravitational waves will set an upper bound on the slope $|\Ls'/\Ls|$ of the species scale. We have found an explicit expression of this bound in terms of the tensor-to-scalar ratio $r$ (see eq.~\eqref{upperboundmain}) and showed that this can be strengthened depending on the class of inflation model considered. For current and near-future target values of the tensor-to-scalar ratio, we have demonstrated that this bound must be an $\mathcal{O}(1)$ quantity, a result consistent with recent theoretical investigations \cite{vandeHeisteeg:2023ubh,Calderon-Infante:2023ler,vandeHeisteeg:2023dlw,Lust:2023zql}. As an example, the case of brane inflation \cite{Dvali:1998pa,Kachru:2003sx,Kallosh:2022feu,Burgess:2022nbx} can yield a stringent upper value such as $|\Ls'/\Ls|\lesssim 0.5$, for percentage values of $r$. These main results, together with a summarizing plot, are presented in sec.~\ref{sec:bounds}.

The existence of an upper bound on the decay rate of the species scale $\Ls$ is intrinsically related to the existence on an infrared cut-off in the EFT \cite{Calderon-Infante:2023ler}. In the context of this work, this situation becomes directly manifest: a detection of primordial tensor perturbations fixes immediately both an upper bound for $|\Ls'/\Ls|$ (main result of this work) and also the scale of inflation $H$ (a famous result that the amplitude of tensor perturbations $A_t\sim H^2$), which acts in fact as infrared cut-off for the effective theory. At the same time, since we have already measured the amplitude of scalar perturbations $A_s$, which is proportional to the ratio $H^2/\epsilon$, a detection of $r$ will give us precious information also about the slope $\epsilon=(V'/V)^2/2$ of the scalar potential $V(\phi)$. Therefore, we have showed that a detection of primordial gravitational waves would connect IR an UV properties of inflationary EFTs. 

Some final comments are in order. First, we would like to note that we have assumed that the species scale varies exponentially in field space. As explained in sec.~\ref{sec:inflation}, one important motivation for this assumption is given by the Lyth bound eq.~\eqref{Lyth}, which suggests that, for any target of current and next-generation experiments, a detection of $r$ would generically indicate a super-Planckian scalar field displacement. In this case, the Swampland Distance Conjecture would in fact point at an exponential variation of $\Ls$ in terms of the canonical inflaton $\phi$. However, one must also acknowledge that, in concrete string theory settings, the species scale $\Ls$ may exhibit a different functional form in the bulk of the moduli space for $\mathcal{O}(1)$ field displacements \cite{vandeHeisteeg:2023ubh,vandeHeisteeg:2023dlw}. Here we would like to stress that, in the case that $\Lambda_s$ grows slower than an exponential, our results are nonetheless  valid and therefore  still provide an upper bound on the decay rate of the species scale.

The second comment regards the fact that, in order to find the upper bound on $|\Ls'/\Ls|$, we have considered that the exponential behaviour of $\Ls$ kicks in when $\Lambda_0$, as defined in eq.~\eqref{speciesscaledecay}, is equal to its maximum possible value, namely the Planck scale $M_{\rm P}$. However it might well be that the exponential behaviour kicks in when the species scale is already some orders below the Planck mass and, therefore, $\Lambda_0<M_{\rm P}$. This situation would lead to a tighter constraint on the decay rate.

In conclusion, we have presented additional evidence showcasing the potential of precision cosmology to provide tangible insights into properties of quantum gravity. The Swampland program has offered a convenient framework and effective tools to establish such a connection and extract concrete results. We look forward to addressing many other interesting open questions in the near future.

\vspace{-0.2cm}

\acknowledgments
\vspace{-0.2cm}
\noindent
We would like to thank Alvaro Herr\'aez, Dieter L\"ust, \mbox{Giada} Maugeri, Carmine Montella for helpful comments.


\bibliographystyle{utphys}
\bibliography{refs}

\end{document}